\documentclass[fleqn,usenatbib]{mnras}
\usepackage{newtxtext,newtxmath}

\usepackage[T1]{fontenc}
\usepackage{ae,aecompl}
\usepackage{graphicx}	
\usepackage{amsmath}	
\usepackage{amssymb}	
\usepackage{epsfig}

\newcommand{\be}{\begin{eqnarray}}
\newcommand{\ee}{\end{eqnarray}}

\newcommand\plottwo[2]{%
 \centering 
 \leavevmode 
 \columnwidth=.45\columnwidth 
 \includegraphics[width={\eps@scaling\columnwidth}]{#1}%
 \hfil 
 \includegraphics[width={\eps@scaling\columnwidth}]{#2}%
}%

\title[Nebular H-$\alpha$ Emission]{H$\alpha$ emission in the nebular spectrum of the Type Ia supernova ASASSN-18tb\thanks{This paper includes data gathered with the 6.5 meter Magellan Telescopes located at Las Campanas Observatory, Chile.}}

\author[Kollmeier et al.]{
Juna A. Kollmeier,$^{1}$\thanks{E-mail: jak@carnegiescience.edu} 
Ping Chen,$^{2}$ 
Subo Dong,$^{2}$
Nidia Morrell,$^{3}$
M. M. Phillips,$^{3}$ 
\newauthor
Doron Kushnir,$^{4}$ 
J. L. Prieto,$^{5,6}$
Anthony L. Piro,$^{1}$ and
Joshua D. Simon$^{1}$
\newauthor
\\
$^{1}${Observatories of the Carnegie Institution for Science, 813 Santa Barbara Street, Pasadena, CA 91101, USA}\\
$^{2}${Kavli Institute for Astronomy and Astrophysics, Peking University, Yi He Yuan Road 5, Hai Dian District, Beijing 100871, China}\\
$^{3}${Las Campanas Observatory, Carnegie Observatories, Casilla
601, La Serena, Chile}\\
$^{4}${Department of Particle Physics and Astrophysics, Weizmann Institute of Science, Rehovot 76100, Israel}\\
$^{5}${N\'ucleo de Astronom\'ia de la Facultad de Ingenier\'ia y Ciencias, Universidad Diego Portales, Av. Ej\'ercito 441, Santiago, Chile}\\
$^{6}${Millennium Institute of Astrophysics, Santiago, Chile}\\
}

\date{Accepted XXX. Received YYY; in original form ZZZ}

\pubyear{2019}

\begin{document}
\label{firstpage}
\pagerange{\pageref{firstpage}--\pageref{lastpage}}

\maketitle

\begin{abstract}

As part of the 100IAS survey, a program aimed to obtain nebular-phase spectra for a volume-limited and homogeneous sample of Type Ia supernovae (SNe~Ia), we observed ASASSN-18tb (SN 2018fhw) at 139 days past maximum light. ASASSN-18tb was a fast-declining, sub-luminous event that
fits well within the observed photometric and spectroscopic distributions of the SN Ia population.  We detect a prominent H$\alpha$ emission line ($L_{{\rm H}\alpha}=2.2\pm0.2\times10^{38}$\,ergs\,s$^{-1}$) with FWHM $\approx1100$\,km\,s$^{-1}$ in the nebular-phase spectrum of this SN Ia.  
High luminosity H$\alpha$ emission ($L_{{\rm H}\alpha}\gtrsim 10^{40}$\,ergs~s$^{-1}$) has previously been discovered in a rare class of SNe Ia-like objects showing CSM interactions (SNe Ia-CSM). They predominantly belong to over-luminous ($M_{\rm max}<-19$\,mag in optical) 1991T-like SNe Ia and are exclusively found in star-forming galaxies. By contrast, ASASSN-18tb is a sub-luminous SN Ia ($M_{B, {\rm max}}\sim -17.7$\,mag) found in an early-type galaxy dominated by old stellar populations.  We discuss possible origins for the observed hydrogen.
Out of 75 SNe Ia for which we have so far obtained nebular spectra in 100IAS, no other SN shows a $\sim 1000 \,{\rm km\,s^{-1}}$  H$\alpha$ emission line with comparable line luminosity as ASASSN-18tb, emphasizing the rarity of such emission in the nebular phase.  Based on preliminary results from our survey, the rate for ASASSN-18tb-like nebular H$\alpha$ emission could be as high as $\sim 10\%$ level among sub-luminous SNe Ia.

\end{abstract}

\begin{keywords}
supernovae: general --
supernovae: individual (ASASSN-18tb/SN2018fhw) --
techniques: spectroscopic
\end{keywords}

\section{Introduction}

\label{sec:Introduction}
Although Type Ia supernovae (SNe~Ia) are fundamentally important to many areas of astrophysics and cosmology, the explosion mechanism and nature of their progenitors remains elusive. It is generally agreed upon that SNe~Ia result from the thermonuclear explosions of C/O white dwarfs \citep[WDs,][]{Hoyle60, nugent11}, but there are unsolved questions about which systems this occurs in and how the explosions proceed. Broadly speaking, the progenitor scenarios can be grouped into the single degenerate (SD) and double degenerate (DD) classes \citep[e.g., see reviews by][]{maoz,wang18}, but even among these there are fundamental differences between the models \citep[e.g., see a review by][]{livio18}. 

In the ``classic" SD scenario, the WD gains material from a non-degenerate companion, such as a main-sequence, sub-giant, or red giant star.
A chief prediction of these models is the stripping of $\sim0.1-0.5\,M_\odot$ of material from the donor \citep{Marietta00,Pan12,Boehner17}, which should be revealed at late times as the ejecta expands and becomes optically thin \citep{mattila05,Botyanszki18}. This basic expectation has motivated extensive observational efforts to find spectral evidence of companion material in the nebular phases of SNe Ia, notably hydrogen emission \citep{mattila05,leonard07,lundqvist13,Lundqvist15,maguire,graham17,Sand18,Shappee18,Holmbo18,Tucker18,Dimitriadis19}, which have so far not detected the predicted signature.  

By definition, SNe Ia {\it lack} signs of hydrogen in the ejecta, but should the explosion occur in a hydrogen-rich medium that, too, would reveal itself in nebular hydrogen emission.  Indeed, there does exist a rare class of Ia-like objects that exhibit evidence for circumstellar interactions (the prototype being SN~2002ic, studied first by \citealt{hamuy03, wood-vasey04}).  The spectra of these so-called ``SNe~Ia-CSM'' are often dominated by highly luminous ($L_{{\rm H}\alpha}\sim 10^{40}-10^{41}$\,ergs~s$^{-1}$) H$\alpha$ emission lines at FWHM $\sim 500-2000$\,km\,s$^{-1}$ \citep{silverman13}. Most of the known SNe~Ia-CSM, however, show H$\alpha$ lines even in the very early phases of spectral evolution, although two of them, PTF11kx \citep{dilday12} and SN 2015cp  \citep{Graham18}, show a delayed CSM interaction. One speculation is that the progenitors of the SNe~Ia-CSM are binary systems with a C/O WD and an asymptotic giant branch star \citep{hamuy03} such as the symbiotic nova RS~Oph \citep{dilday12}.  Whether these are interactions with a dense interstellar medium, the long-sought ``smoking gun" of the classic SD models, or evidence for alternate progenitor models remains uncertain.  What is clear is that, for a variety of reasons, we might expect hydrogen emission in the late-time nebular spectra for SNe Ia, but it has been difficult to find.

Here we report nebular-phase spectroscopy of ASASSN-18tb, where we find clear evidence of hydrogen in emission. This observation was obtained as part of the 100 type IA Supernova (``100IAS'') survey, where we aim to systematically obtain nebular spectra of 100 SNe~Ia from a volume-limited sample of nearby SNe Ia  \citep{dong18}. In \S~\ref{sec:obs}, we describe our observations and data analysis. In \S~\ref{sec:csm}, we discuss possible interpretations and implications of this result. We conclude in \S~\ref{sec:conclusion} with a discussion of future work.

\section{Observations and Data Analysis}
\label{sec:obs}
\begin{figure}\label{branch}
    \centering
    \includegraphics[width=0.45\textwidth,trim=1.0cm 1.5cm 1.0cm 0.5cm]{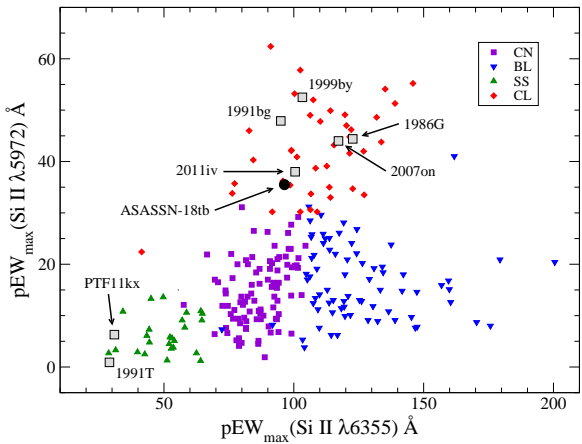}
    \caption{Branch Diagram \citep{Branch06} placing ASASSN-18tb in context of the general SN Ia population.  The Branch spectroscopic classifications are ``Core Normal''(CN), ``Cool'' (CL), ``Broad Line'' (BL), and ``Shallow Silicon'' (SS) based on pseudo equivalent width (pEW) measurements of the Si II $\lambda$6355\,\AA\,and Si II $\lambda$5972\,\AA\,lines. We show for comparison the location of a few well-observed sub-luminous SNe Ia that belong to the CL subclass. Although near the ``borderline'' between CL and CN events, ASASSN-18tb clearly belongs to CL subclass. Also plotted are points corresponding to the SS event SN~1991T and the SN~Ia-CSM PTF11kx. The data in this diagram are from \citet{blondin12} and \citet{gall18}, except for PTF11kx and ASASSN-18tb, whose spectra were measured for this paper.}
    \label{fig:branch}
\end{figure}

ASASSN-18tb (SN~2018fhw) was discovered by the All-Sky Automated Survey for Supernovae (ASAS-SN, \citealt{asassn} \citealt{atel1}) and was classified as a SN Ia based on a SALT spectrum taken on UT 2018-08-23 \citep{atel2}. The publicly-available spectrum of the SN on the Transient Name Server (https://wis-tns.weizmann.ac.il/object/2018fhw) was obtained at a phase of $-4$~days with respect to the epoch of maximum light in the $B$ band.  We show in Figure~\ref{fig:branch} a Branch Diagram \citep{Branch06} to place ASASSN-18tb in context with other SNe.  Pseudo equivalent widths of the \ion{Si}{II} $\lambda$5972 and $\lambda$6355 absorption features measured from the SALT spectrum and corrected to maximum light using the relations given in Table~7 of \citet{folatelli13} indicate that ASASSN-18tb was of the ``Cool'' (CL) subtype in the Branch classification system.

\begin{figure*}
    \centering
    {\includegraphics[width=0.5\textwidth,trim=0.0cm 1.5cm 0.0cm 0.5cm]{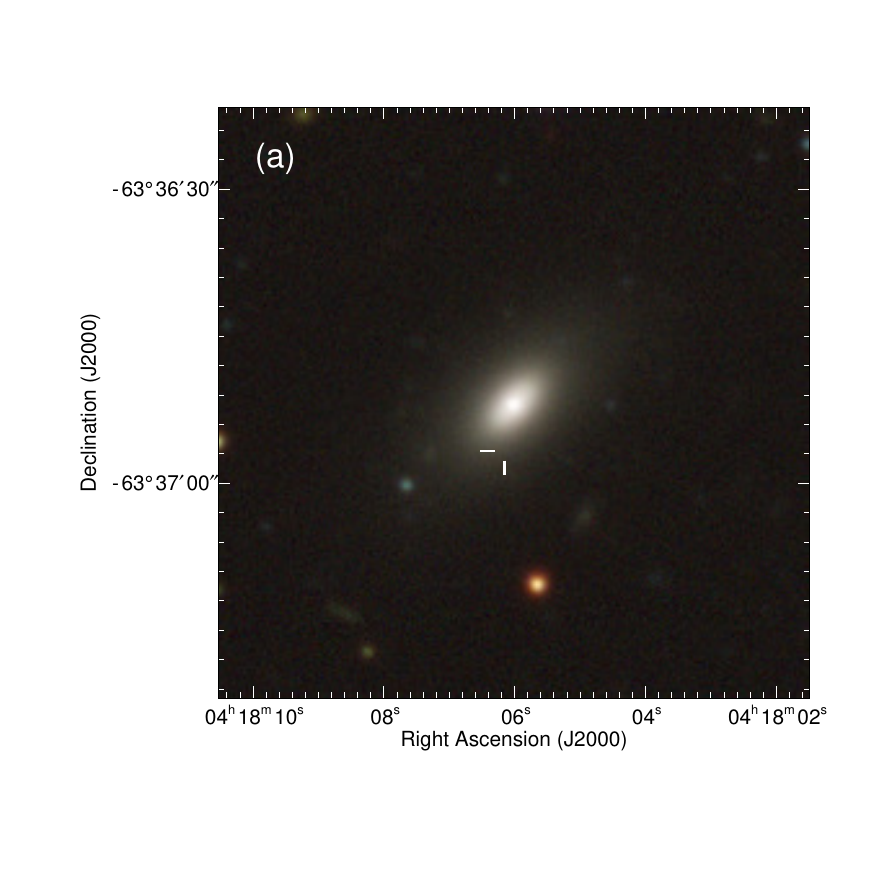}}\hfill
   {\includegraphics[width=0.5\textwidth,trim=0.0cm 1.5cm 0.0cm 0.5cm]{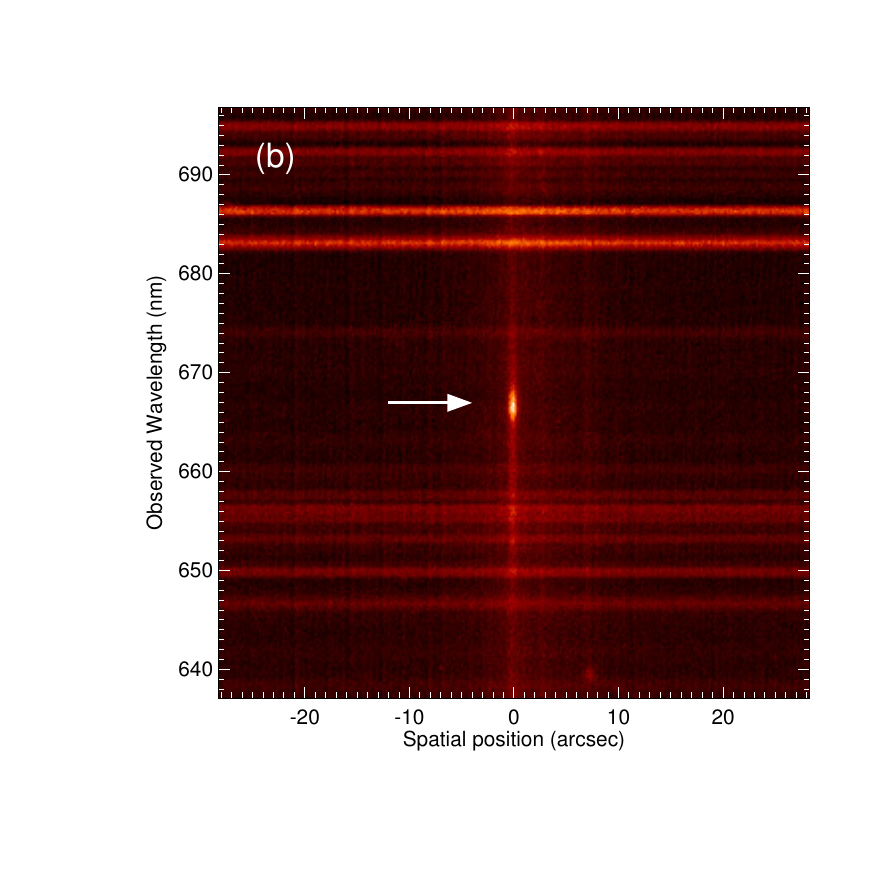}}
    \caption{(a) gri image of the host galaxy of ASASSN-18tb, 2MASX J04180598-6336523, taken from Data Release 1 of the Dark Energy Survey \citep{desdr1}.  The image is 1\arcmin\ on a side (21 kpc at the distance of the SN) and the position of the SN is marked by the white ticks.  (b) Two-dimensional Magellan/LDSS3 spectrum of ASASSN-18tb near the wavelength of H$\alpha$.  The narrow H$\alpha$ emission line we detected is indicated with an arrow. The faint emission observed at $\lambda \sim 640$~nm and $\sim$7~arcsec to the right of the SN is due to [O~II]$\lambda 3727$ from an unrelated faint galaxy at $z \sim 0.7$.}
    \label{fig:2Dspec}
\end{figure*}

Light curves in $BVri$ obtained at the Las Cumbres Observatory were fit using the ``SNooPy'' \citep{burns11} package to yield a time of $B$ maximum of JD~$2458357.3 \pm 0.4$, a maximum-light magnitude of $B = 16.66 \pm 0.04$, and decline rate parameters of $\Delta{\rm m}_{15}(B) = 2.0 \pm 0.1$ \citep{phillips93} and $s_{BV} = 0.5 \pm 0.04$ \citep{burns14}.  The absolute $B$ magnitude at maximum (assuming zero host galaxy reddening and $H\rm{_0 = 72~km~s^{-1}~Mpc^{-1}}$) was $-17.7 \pm 0.06$~mag, placing ASASSN-18tb on the sub-luminous portion of the $s_{BV}$ vs. $M(B_{max})$ relationship for SNe~Ia \citep[see Figure 4 of][]{burns18}. Shown in Figure~\ref{fig:2Dspec}a is an image of the host galaxy, 2MASX J04180598-6336523, an apparent dwarf elliptical galaxy ($M_{g} = -17.4$) at $z_{helio} = 0.0170\pm0.0001$ \citep{atel2}.  The $g-r$ and $r-i$ colors of the host measured by the Dark Energy Survey \citep{desdr1} are consistent with a dwarf elliptical \citep{schombert18}, and our optical spectrum shows no obvious emission lines. We note that this classification is in line with the well-known association of sub-luminous SNe~Ia with early-type galaxies \citep{hamuy96,hamuy00}.

The nebular spectrum presented here was obtained with LDSS3 (Low Dispersion Survey Spectrograph 3) on UT 2019-01-13 at the 6.5m Magellan Clay telescope as part of 100IAS. The post-maximum phase at the time of observation was  $\approx +139$\,days. We obtained three 1800~s exposures of ASASSN-18tb with spectral resolving power $\rm{R}\approx700$.  We had sub-arcsecond seeing throughout our exposure sequence with the final seeing settling to 0.65\arcsec.  To avoid as much host galaxy contamination of the spectrum as possible, instead of aligning the slit along the parallactic angle (which had an average value of 24$^\circ$ during the observations), we rotated it to a position angle of $90.5^\circ$.  We employed routine calibration and reduction procedures using \textsc{iraf} tasks.

\begin{figure}\label{spectrum}
    \centering
    \includegraphics[width=0.45\textwidth,trim=1.0cm 5.5cm 1.0cm 3.0cm]{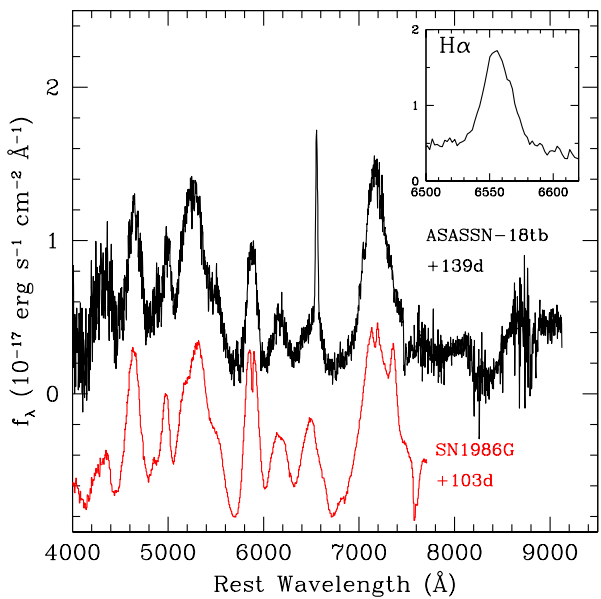}
    \caption{The rest-frame nebular-phase spectrum of ASASSN-18tb. The full $\approx 139$\,days Magellan spectrum is shown in black. The inset shows a zoom-in on the strong H$\alpha$ 6562.8 \AA\ line detection. We also show a spectrum (scaled and shifted in flux) of the sub-luminous Type~Ia SN~1986G at 103~days in red for comparison.}
    \label{fig:1Dspec}
\end{figure}

We show the flat-fielded, rectified, and wavelength-calibrated two-dimensional spectrum in Figure~\ref{fig:2Dspec}b, where the H$\alpha$ line is clearly apparent (and indicated with an arrow).  Note that in the 2D spectrum shown in Figure~\ref{fig:2Dspec}b there is no evidence of spatially extended narrow H$\alpha$ emission, indicating that the SN exploded in a region of low star formation. The extracted 1D spectrum of ASASSN-18tb is displayed in Figure~\ref{fig:1Dspec}, with an inset showing the strong H$\alpha$ emission line detection. We also include the spectrum (scaled and shifted in flux) of the sub-luminous Type~Ia SN~1986G \citep{phillips87} obtained at 103~days after $B$-band maximum \citep{cristiani92}, showing similar features and line ratios compared to the nebular-phase spectrum of ASASSN-18tb.  

The emission line profile of the H$\alpha$ feature is reasonably well fit by a Gaussian with a FWHM of 26~\AA.  Taking into account the resolution of the spectrum and the redshift of the host galaxy, this corresponds to a FWHM of $1085~{\rm km~s}^{-1}$.  The peak of the emission is blueshifted by $\sim 300~{\rm km~s}^{-1}$ with respect to the redshift of the host galaxy, and the total measured flux is $3.5 \times 10^{-16}~{\rm ergs~s^{-1}~cm~^{-2}}$, with an estimated uncertainty of $\pm 10$\%.  Assuming a distance of 70.7~Mpc implies a luminosity of $2.2 \pm 0.2 \times 10^{38}~{\rm ergs~s^{-1}}$.

No other emission features with similar width as the H$\alpha$ line are detectable in our spectrum of ASASSN-18tb.  In particular, no obvious emission is observed at the expected position of H$\beta$.  From Equation 1 of \citet{leonard07}, we calculate a 3$\sigma$ upper limit for the equivalent width of H$\beta$ emission of 3.4~\AA.  Given the continuum flux at the wavelength of H$\beta$ after correcting for slit losses from differential atmospheric refraction (following \citealt{shappee17}), this value translates to an upper limit to the flux of H$\beta$ of $2.6 \times 10^{-17}~{\rm ergs~s^{-1}~cm~^{-2}}$, and $F({\rm H}\alpha) / F({\rm H}\beta) \gtrsim 14$.  However, visually comparing a Gaussian emission line with this flux to the observed spectrum at H$\beta$ suggests that such a feature would not, in fact, be detected at 3$\sigma$ significance, and this limit is therefore insufficiently conservative.  By shifting and scaling the observed H$\alpha$ profile to the wavelength of H$\beta$, we estimate that a more realistic lower limit on the Balmer decrement is $F({\rm H}\alpha) / F({\rm H}\beta) \gtrsim 6$. We further discuss this Balmer decrement in relation to the origin of this hydrogen emission next.
\footnote{We note that while the observed limit on the Balmer decrement is not uncommon in astrophysical plasmas, the deviation from the standard recombination value requires explanation, e.g. dust or an alternate power source.}

\section{Discussion}
\label{sec:csm}

The search for hydrogen emission in SNe~Ia has been ongoing for many decades.  As mentioned above, the presence of hydrogen and/or helium-rich material in the CSM is regarded as a generic prediction of models that favor single degenerate binary systems as SNe~Ia progenitors.  Hydrogen stripped from a non-degenerate companion embedded in the SN ejecta at velocities of $\sim$1000--2000~${\rm km~s}^{-1}$ is expected to be observable at epochs $\gtrsim$~200~days after explosion \citep[see][and references therein]{Botyanszki18}. However, the search has proven elusive, with non-detections being the rule (e.g. \citealt{mattila05, leonard07, lundqvist13, shappee13, Shappee18}). A single tentative detection of H$\alpha$ emission for SN 2013ct has been reported by \citet{maguire}, who also reported 17 non-detections.  In the remainder of this section, we discuss possible scenarios for explaining the H$\alpha$ emission in ASASSN-18tb.

\subsection{Comparison with SNe Ia-CSM Objects}
Comparison of ASASSN-18tb with the class of ``SNe Ia-CSM" is
instructive. At maximum light, the spectrum of SN~2002ic (the prototype Ia-CSM) showed absorption features typical of a luminous, slow-declining 1991T-like event \citep{phillips92, filippenko92}, but diluted in strength by the CSM interaction with the SN ejecta.  Since then, several further examples of SNe~Ia-CSM objects have been identified \citep{silverman13}, the best-observed one being PTF11kx, which showed temporally variable \ion{Na}{I}, \ion{Fe}{II}, \ion{Ti}{II}, and \ion{He}{I} absorption lines and a 1991T-like spectrum at maximum before developing strong hydrogen and calcium emission lines approximately a month after maximum which was interpreted as due to the interaction of the SN ejecta with a shell of CSM at a distance of $\sim10^{16}$\,cm \citep{dilday12}.  Recently, the detection of hydrogen and calcium emission 664~days after maximum in the 1991T-like SN~2015cp has provided a second case of a delayed CSM interaction in a SN~Ia-CSM event \citep{Graham18}. 

Our detection of H$\alpha$ emission in the spectrum of the sub-luminous SN~Ia ASASSN-18tb adds a new element to the puzzle. The Balmer decrement we observe is similar to the SNe~Ia-CSM, which is consistent with what might be expected from shock excitation. But how do the spectral and photometric properties of ASASSN-18tb compare to the SNe~Ia-CSM? The early epochs of the best-observed examples of the SNe~Ia-CSM resemble those of the luminous, slow-declining SN~1991T.  Figure~\ref{fig:branch} shows the positions of SN~1991T and the SN~Ia-CSM PTF11kx in the Branch diagram.  These objects are members of the ``Shallow-Silicon'' (SS) subtype, and thus are quite different in their spectral and photometric characteristics compared to the fast-declining, sub-luminous ASASSN-18tb which belongs to the ``Cool'' (CL) region of the Branch diagram.  Furthermore, all previous SNe Ia-CSM were found in star-forming galaxies, while the host of ASASSN-18tb is an early-type galaxy dominated by old stellar populations.  Finally, the H$\alpha$ luminosity of ASASSN-18tb is two orders of magnitude smaller than the H$\alpha$ luminosities of SNe~Ia-CSM (Silverman et al. 2013; Graham et al. 2018).

The evidence is thus mixed as to whether ASASSN-18tb should be considered similar to the SN~Ia-CSM objects or not.  It is possible that ASASSN-18tb reveals a much broader variety within the SNe~Ia-CSM class.  It is also possible that ASASSN-18tb is something distinctive altogether.  Definitively answering this question will require follow up observations, which we discuss in further detail below.

\subsection{Stripped Mass Estimates in the SD Model}
As described above, broad ($\sim1000$\,km\,s$^{-1}$), nebular-phase H$\alpha$ emission due to stripped material from a non-degenerate companion is a longstanding prediction of SD models \citep[e.g.][]{mattila05,Botyanszki18}.  While the full multi-dimensional radiation transfer calculations required to model this process accurately do not exist in the literature, for illustrative purposes, we simply translate the H$\alpha$ luminosity of ASASSN-18tb into a stripped hydrogen mass according to a series of  MS38 models \citep{B17} with reduced density using the simplified radiation transfer modeling by \citet{Botyanszki18}.  We obtain $\sim2\times10^{-3}\,M_\odot$. This value is significantly lower than the $\gtrsim 0.1\,M_\odot$ typically expected from SD models, and whether much lower stripped masses are  possible (e.g., at a large binary separation) is under debate \citep{Liu12,B17}.  It is curious, however, that the ratio of the peak flux of H$\alpha$ to that of the [FeIII]$\lambda$4658 feature in the nebular spectrum observed in ASASSN-18tb is similar to that predicted by \citet[][see their Figure 2]{Botyanszki18}, despite the fact that our mass estimate is considerably lower than their prediction.  Note that the Botyanszki et al. calculations are for a Chandrasekhar-mass WD with 0.6 solar masses of $^{56}$Ni, whereas ASASSN18-tb was a sub-luminous event, possibly sub-Chandrasekhar-mass WD \citep[e.g.,][]{Mazzali97,Piro14,Scalzo14,Wygoda19}, with a Ni mass likely in the range of $\sim0.1-0.2\,M_\odot$.  Accounting for the lower Ni mass is likely to increase the stripped-mass estimate to $\sim 0.01 M_\odot$.  We look forward to future models more appropriate to sub-luminous SNe.

A separate and distinct prediction of the SD scenario is the appearance of excess radiation during the first $\sim1-2$ days following explosion due to the interaction of the SN ejecta with its non-degenerate companion \citep{Kasen10}.  Fortuitously, when ASASSN-18tb was discovered, it was located within the Camera 4 field of the Transiting Exoplanet Survey Satellite (TESS; \citealt{tess}) Sector 1 observations \citep{atel1}. The precise TESS light curves from this SN may provide an important probe of these models, such as whether such a collision occurs or if there is additional emission at early times from dense CSM interaction \citep{Piro16}.

\subsection{Other Possibilities}
Finally, we raise the possibility that the observed H$\alpha$ has an entirely different origin.  One alternative explanation is that the H$\alpha$ emission is the result of fluorescent UV pumping.  This would naturally produce a high Balmer decrement (as observed).  The cartoon picture is that of an expanding hydrogen shell being pumped by the soft UV photons from the SNe itself.  In general, weak Paschen lines would indicate a non-recombination source and the H$\alpha$ flux would scale with the UV source flux.  This highlights the important role UV and IR observations play in understanding these systems.  

Given the ubiquity of hydrogen in the universe and plethora of mechanisms to power H$\alpha$ (e.g., stellar winds/mass losses, common envelope/Roche lobes, planetary nebulae, interstellar media interactions, shock interactions, to name a few), there are many other possible scenarios that we have not discussed to interpret this rare discovery. Indeed, it is perhaps the persistent {\it rarity} of finding nebular hydrogen emission in these systems that may be the biggest clue to solving the Type Ia progenitor puzzle.

\subsection{Statistical Preview}
\begin{figure}\label{spectra}
    \centering
    \includegraphics[width=0.5\textwidth,trim=0.5cm 1.0cm 0.0cm 1.0cm]{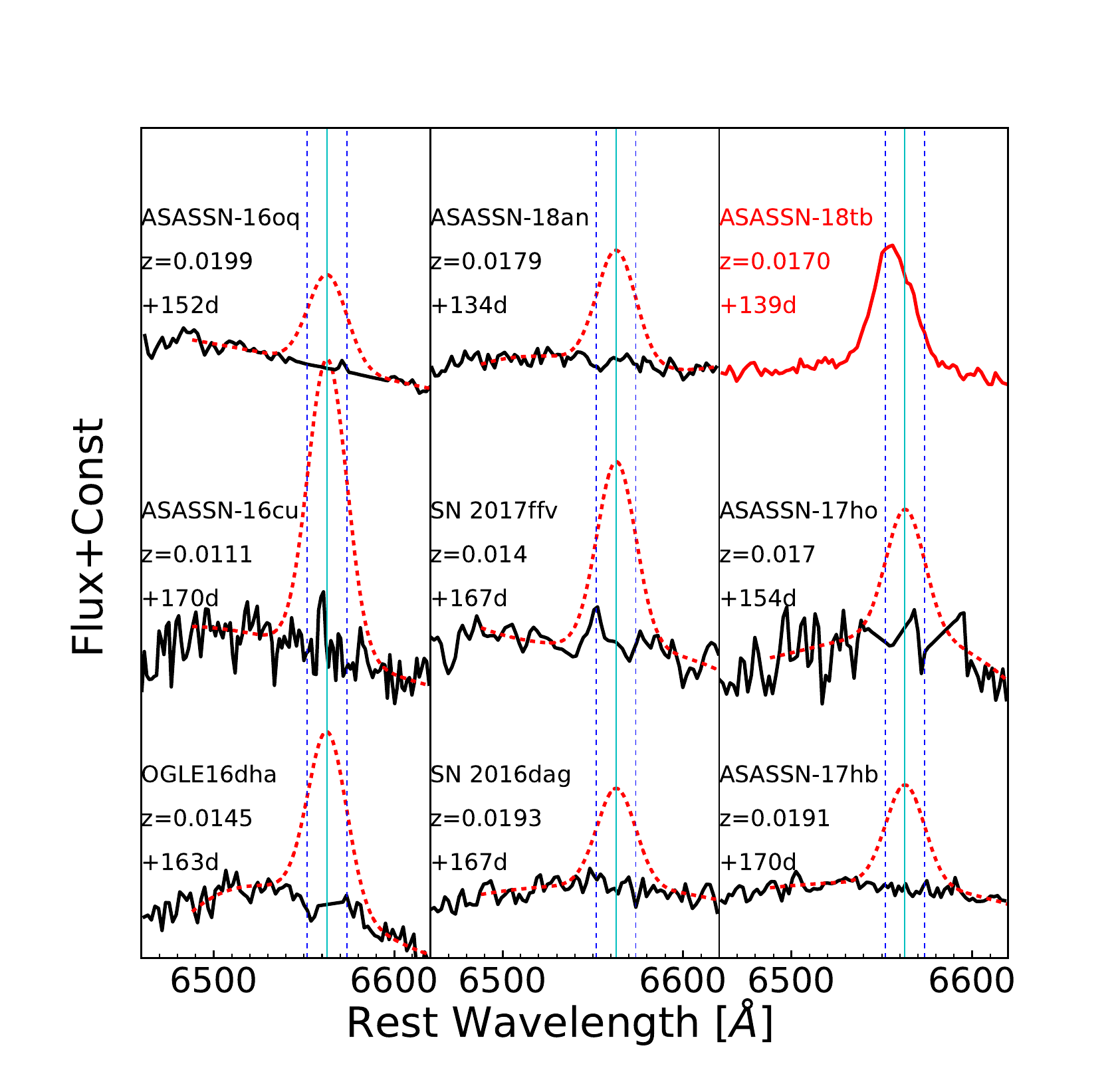}
    \caption{A sample of nebular spectra from the 75 SNe Ia of the 100IAS survey. Except for ASASSN-18tb (red; upper right), no other object in 100IAS show FWHM$\sim 1000$\,km\,s$^{-1}$ H$\alpha$ detected at $\geq 2\times10^{38}$\,erg\,s$^{-1}$, shown as simulated signals with red dashed lines. The rest-frame wavelength of H$\alpha$ is shown by the solid cyan lines and at $\pm 500$\,km\,s$^{-1}$ by blue dashed lines in each sub panel.}
    \label{fig:fig4}
\end{figure}

The 100IAS survey was constructed precisely for the purpose of obtaining a complete, nearly volume-limited sample of TypeIa SNe, in order to gauge nebular-phase demographics for these objects.  By construction, we sample broadly across the Type Ia luminosity function, and notably, we aim to be unbiased with respect to sub-luminous events.  Over the course of 100IAS, we have thus far obtained nebular-phase spectra for 75 SNe Ia.  For all objects except ASASSN-18tb, H$\alpha$ emission with a FWHM$\approx 1100~{\rm km~s}^{-1}$ and a luminosity of $2\times10^{38}$\,ergs~s$^{-1}$ can be confidently ruled out. A few examples are shown in Figure~\ref{fig:fig4}. Therefore, the luminous H$\alpha$ emission seen in ASASSN-18tb must be rare, with an occurrence rate of $\sim$~few\,\% of all SNe Ia. Its rarity is also consistent with the nondetections of such H$\alpha$ emission lines for several dozen SNe~Ia with well-observed nebular-phase spectra published in the literature (see references in Section 1).
With a peak luminosity of $M_{B,\rm max} =-17.7 \pm 0.06$~mag, ASSASN-18tb is at the low-luminosity end of the SNe Ia luminosity function (see, e.g., \citealt{burns14}), and nebular spectra for sub-luminous SNe Ia are dramatically under-represented in the literature. In the 100IAS sample, there are $\sim 10$ objects with comparable or lower luminosity to ASASSN-18tb, and so the fraction of SNe~Ia with luminous H$\alpha$ at nebular phase for this population could be as high as $\sim 10\%$.

\section{Conclusions}
\label{sec:conclusion}

We present an unambiguous detection of H$\alpha$ emission in the nebular-phase spectrum of the Type Ia SNe ASASSN-18tb as part of the 100IAS survey. We find some similarities with the rare class of SNe~Ia-CSM, we also highlight some striking differences, such as the photometric and spectral properties of ASASSN-18tb and the relatively low luminosity of its H$\alpha$ emission in comparison to this class.  With our data alone, we cannot definitively rule out a CSM-interaction interpretation for this feature as opposed to alternate hypotheses such as the long-sought stripped material from a non-degenerate companion, or other potential scenarios with which we have not compared. In any case, ASASSN-18tb is worthy of significant follow up.

Such follow up observations will be key for discriminating between the possible origins of the H$\alpha$ emission. For example, the UV/X-ray emission may prove critical. This has indeed been observed for at least one SN~Ia-CSM \citep{Bochenek18}, although we note that this is one of the most extreme members of this class. CSM interaction also results in a larger bolometric luminosity that declines more shallowly than cobalt decay \citep{silverman13}, and infrared observations should be sought to construct a late time light curve. Further spectra as the SN fades may reveal additional emission lines as the shock interaction increases. We look forward to further comprehensive theoretical and observational analysis of this system to decode the physical origin of this H$\alpha$ emission and thereby gain further insights into the elusive progenitors of SNe Ia.

\section*{Acknowledgements}

We thank Boaz Katz for important discussions and insights as well as significant contributions to this paper.  P.C. and S.D. acknowledge Project 11573003 supported by NSFC. 
This project used public archival data from the Dark Energy Survey (DES). Funding for the DES Projects has been provided by the U.S. Department of Energy, the U.S. National Science Foundation, the Ministry of Science and Education of Spain, the Science and Technology Facilities Council of the United Kingdom, the Higher Education Funding Council for England, the National Center for Supercomputing Applications at the University of Illinois at Urbana-Champaign, the Kavli Institute of Cosmological Physics at the University of Chicago, the Center for Cosmology and Astro-Particle Physics at the Ohio State University, the Mitchell Institute for Fundamental Physics and Astronomy at Texas A\&M University, Financiadora de Estudos e Projetos, Funda{\c c}{\~a}o Carlos Chagas Filho de Amparo {\`a} Pesquisa do Estado do Rio de Janeiro, Conselho Nacional de Desenvolvimento Cient{\'i}fico e Tecnol{\'o}gico and the Minist{\'e}rio da Ci{\^e}ncia, Tecnologia e Inova{\c c}{\~a}o, the Deutsche Forschungsgemeinschaft, and the Collaborating Institutions in the Dark Energy Survey. The Collaborating Institutions are Argonne National Laboratory, the University of California at Santa Cruz, the University of Cambridge, Centro de Investigaciones Energ{\'e}ticas, Medioambientales y Tecnol{\'o}gicas-Madrid, the University of Chicago, University College London, the DES-Brazil Consortium, the University of Edinburgh, the Eidgen{\"o}ssische Technische Hochschule (ETH) Z{\"u}rich,  Fermi National Accelerator Laboratory, the University of Illinois at Urbana-Champaign, the Institut de Ci{\`e}ncies de l'Espai (IEEC/CSIC), the Institut de F{\'i}sica d'Altes Energies, Lawrence Berkeley National Laboratory, the Ludwig-Maximilians Universit{\"a}t M{\"u}nchen and the associated Excellence Cluster Universe, the University of Michigan, the National Optical Astronomy Observatory, the University of Nottingham, The Ohio State University, the OzDES Membership Consortium, the University of Pennsylvania, the University of Portsmouth, SLAC National Accelerator Laboratory, Stanford University, the University of Sussex, and Texas A\&M University.  This research uses data obtained through the Telescope Access Program (TAP), which has been funded by the National Astronomical Observatories of China, the Chinese Academy of Sciences, and the Special Fund for Astronomy from the Ministry of Finance. Based in part on observations at Cerro Tololo Inter-American Observatory, National Optical Astronomy Observatory, which is operated by the Association of Universities for Research in Astronomy (AURA) under a cooperative agreement with the National Science Foundation.

\bsp	
\label{lastpage}

\begin{thebibliography}{99}
\bibitem[Abbott et al.(2018)]{desdr1} Abbott, T.~M.~C., Abdalla, F.~B., Allam, S., et al.\ 2018, \apjs, 239, 18 



\bibitem[Blondin et al.(2012)]{blondin12} Blondin, S., Matheson, T., Kirshner, R.~P., et al.\ 2012, \aj, 143, 126

\bibitem[Bochenek et al.(2018)]{Bochenek18} Bochenek, C.~D., Dwarkadas, V.~V., Silverman, J.~M., et al.\ 2018, \mnras, 473, 336 

\bibitem[Boehner et al.(2017)]{Boehner17} Boehner, P., Plewa, T., \& Langer, N.\ 2017, \mnras, 465, 2060 


\bibitem[Boty{\'a}nszki \& Kasen(2017)]{B17} Boty{\'a}nszki, J., \& Kasen, D.\ 2017, \apj, 845, 176 

\bibitem[Boty{\'a}nszki et al.(2018)]{Botyanszki18} Boty{\'a}nszki, J., Kasen, D., \& Plewa, T.\ 2018, \apjl, 852, L6

\bibitem[Branch et al.(2006)]{Branch06} Branch, D., Dang, L.~C., Hall, N., et al.\ 2006, \pasp, 118, 560

\bibitem[Brimacombe et al.(2018)]{atel1} Brimacombe, J. et al..\ 2018, The Astronomer's Telegram, 11976

\bibitem[Burns et al.(2011)]{burns11} Burns, C., et al. 2011, \aj, 141, 19

\bibitem[Burns et al.(2014)]{burns14}  Burns, C.~R., Stritzinger, M., Phillips, M.~M., et al.\ 2014, \apj, 789, 32

\bibitem[Burns et al.(2018)]{burns18} Burns, C.~R., Parent, E., Phillips, M.~M., et al.\ 2018, \apj, 869, 56

\bibitem[Cristiani et al.(1992)]{cristiani92} Cristiani, S., Cappellaro, E., Turatto, M., et al.\ 1992, \aap, 259, 63 

\bibitem[Dilday et al.(2012)]{dilday12} Dilday, B., Howell, D.~A., Cenko, S.~B., et al. 2012, Science, 337, 942

\bibitem[Dimitriadis et al.(2019)]{Dimitriadis19} Dimitriadis, G., Rojas-Bravo, C., Kilpatrick, C.~D., et al.\ 2019, \apjl, 870, L14 

\bibitem[Dong et al.(2018)]{dong18} Dong, S., Katz, B., Kollmeier, J., et al.\ 2018, \mnras, 479, L70

\bibitem[Eweis et al.(2019)]{atel2} 
Eweis, Y. et al.\ 2018, The Astronomer's Telegram, 11980


\bibitem[Filippenko et al.(1992)]{filippenko92} Filippenko, A.~V., Richmond, M.~W., Matheson, T., et al.\ 1992b, \apjl, 384, L15

\bibitem[Folatelli et al.(2013)]{folatelli13} Folatelli, G., Morrell, N., Phillips, M.~M., et al.\ 2013, \apj, 773, 53

\bibitem[Gall et al.(2018)]{gall18} Gall, C., Stritzinger, M.~D., Ashall, C., et al.\ 2018, \aap, 611, A58

\bibitem[Graham et al.(2017)]{graham17} Graham, M.~L., Harris, C.~E., Fox, O.~D., et al.\ 2017, \apj, 843, 102 

\bibitem[Graham et al.(2018)]{Graham18} Graham, M.~L., Harris, C.~E., Nugent, P.~E., et al.\ 2018, arXiv:1812.02757 

\bibitem[Hamuy et al.(1996)]{hamuy96} Hamuy, M., Phillips, M.~M., Suntzeff, N.~B., et al.\ 1996, \aj, 112, 2391

\bibitem[Hamuy et al.(2000)]{hamuy00} Hamuy, M., Trager, S.~C., Pinto, P.~A., et al.\ 2000, \aj, 120, 1479

\bibitem[Hamuy et al.(2003)]{hamuy03} Hamuy, M., Phillips, M.~M., Suntzeff, N.~B., et al. 2003, Nature, 424, 651

\bibitem[Holmbo et al.(2018)]{Holmbo18} Holmbo, S., Stritzinger, M.~D., Shappee, B.~J., et al.\ 2018, arXiv:1809.01359 

\bibitem[Hoyle \& Fowler(1960)]{Hoyle60} Hoyle, F., \& Fowler, W.~A.\ 1960, \apj, 132, 565 

\bibitem[Kasen(2010)]{Kasen10} Kasen, D.\ 2010, \apj, 708, 1025 


\bibitem[Leonard(2007)]{leonard07} Leonard, D.~C.\ 2007, \apj, 670, 1275 


\bibitem[Liu et al.(2012)]{Liu12} Liu, Z.~W., Pakmor, R., R{\"o}pke, F.~K., et al.\ 2012, \aap, 548, A2

\bibitem[Livio \& Mazzali(2018)]{livio18} Livio, M., \& Mazzali, P.\ 2018, \physrep, 736, 1 

\bibitem[Lundqvist et al.(2013)]{lundqvist13} Lundqvist, P., Mattila, S., Sollerman, J., et al.\ 2013, \mnras, 435, 329 

\bibitem[Lundqvist et al.(2015)]{Lundqvist15} Lundqvist, P., Nyholm, A., Taddia, F., et al.\ 2015, \aap, 577, A39 

\bibitem[Maguire et al.(2016)]{maguire} Maguire, K., Taubenberger, S., Sullivan, M., \& Mazzali, P.~A.\ 2016, \mnras, 457, 3254 

\bibitem[Maoz et 
al.(2014)]{maoz} Maoz, D., Mannucci, F., \& Nelemans, G.\ 2014, \araa, 52, 107

\bibitem[Marietta et al.(2000)]{Marietta00} Marietta, E., Burrows, A., \& Fryxell, B.\ 2000, \apjs, 128, 615 

\bibitem[Mattila et al.(2005)]{mattila05} Mattila, S., Lundqvist, P., Sollerman, J., et al.\ 2005, \aap, 443, 649 



\bibitem[Mazzali et al.(1997)]{Mazzali97} Mazzali, P.~A., Chugai, N., Turatto, M., et al.\ 1997, \mnras, 284, 151 

\bibitem[Nugent et al.(2011)]{nugent11} Nugent, P.~E., Sullivan, M., Cenko, S.~B., et al.\ 2011, \nat, 480, 344 



\bibitem[Pan et al.(2012)]{Pan12} Pan, K.-C., Ricker, P.~M., \& Taam, R.~E.\ 2012, \apj, 750, 151 

\bibitem[Phillips(1993)]{phillips93} Phillips, M. M. 1993, \apjl, 413, L105

\bibitem[Phillips et al.(1992)]{phillips92} Phillips, M.~M., Wells, L.~A., Suntzeff, N.~B., et al.\ 1992, \aj, 103, 1632

\bibitem[Phillips et al.(1987)]{phillips87} Phillips, M.~M., Phillips, A.~C., Heathcote, S.~R., et al.\ 1987, \pasp, 99, 592 


\bibitem[Piro et al.(2014)]{Piro14} Piro, A.~L., Thompson, T.~A., \& Kochanek, C.~S.\ 2014, \mnras, 438, 3456 

\bibitem[Piro \& Morozova(2016)]{Piro16} Piro, A.~L., \& Morozova, V.~S.\ 2016, \apj, 826, 96 

\bibitem[Ricker et al.(2014)]{tess} Ricker, G.~R., Winn, J.~N., Vanderspek, R., et al.\ 2014, \procspie, 9143, 914320 

\bibitem[Sand et al.(2018)]{Sand18} Sand, D.~J., Graham, M.~L., Boty{\'a}nszki, J., et al.\ 2018, \apj, 863, 24 

\bibitem[Scalzo et al.(2014)]{Scalzo14} Scalzo, R.~A., Ruiter, A.~J., \& Sim, S.~A.\ 2014, \mnras, 445, 2535 


\bibitem[Schombert(2018)]{schombert18} Schombert, J.~M.\ 2018, \aj, 155, 69


\bibitem[Shappee et al.(2018)]{Shappee18} Shappee, B.~J., Piro, A.~L., Stanek, K.~Z., et al.\ 2018, \apj, 855, 6 

\bibitem[Shappee et al.(2014)]{asassn} Shappee, B.~J., Prieto, J.~L., Grupe, D., et al.\ 2014, \apj, 788, 48

\bibitem[Shappee et al.(2017)]{shappee17} Shappee, B.~J., Simon, J.~D., Drout, M.~R., et al.\ 2017, Science, 358, 1574 

\bibitem[Shappee et al.(2013)]{shappee13} Shappee, B.~J., Stanek, K.~Z., Pogge, R.~W., \& Garnavich, P.~M.\ 2013, \apjl, 762, L5




\bibitem[Silverman et al.(2013)]{silverman13} Silverman, J.~M., Nugent, P.~E., Gal-Yam, A., et al.\ 2013, \apjs, 207, 3



\bibitem[Tucker et al.(2018)]{Tucker18} Tucker, M.~A., Shappee, B.~J., \& Wisniewski, J.~P.\ 2018, arXiv:1811.09635 

\bibitem[Wang(2018)]{wang18} Wang, B.\ 2018, RAA, 18, 49 

\bibitem[Wood-Vasey et al.(2004)]{wood-vasey04} Wood-Vasey, W.~M., Wang, L., \& Aldering, G.\ 2004, \apj, 616, 339

\bibitem[Wygoda et al.(2019)]{Wygoda19} Wygoda, N., Elbaz, Y., \& Katz, B.\ 2019, \mnras


\end{thebibliography}
\end{document}